\documentclass[12pt]{article}

\usepackage{graphicx}
\begin{document}
\begin{center}
{\Large Quantum Gravity Momentum Representation and Maximum Invariant Energy}
\vskip 0.3 true in {\large J. W. Moffat}
\date{} \vskip 0.3 true in
{\it Perimeter Institute for Theoretical Physics,
Waterloo, Ontario N2J2W9, Canada}
\vskip 0.1 true in
and
\vskip 0.1 true in
{\it Department of Physics, University of Toronto, Toronto, Ontario
M5S1A7, Canada}

\end{center}

\begin{abstract}%
We use the idea of the symmetry between the spacetime coordinates $x^\mu$
and the energy-momentum $p^\mu$ in quantum theory to construct a
momentum space quantum gravity geometry with a metric $s_{\mu\nu}$ and a
curvature tensor ${P^\lambda}_{\mu\nu\rho}$. For a closed maximally symmetric
momentum space with a constant 3-curvature, the volume of the p-space admits a
cutoff with an invariant maximum momentum $a$. A Wheeler-DeWitt-type wave equation
is obtained in the momentum space representation. The vacuum energy density and the
self-energy of a charged particle are shown to be finite, and modifications of the
electromagnetic radiation density and the entropy density of a system of particles
occur for high frequencies.

\end{abstract}
\vskip 0.2 true in e-mail: john.moffat@utoronto.ca

\vskip 0.3 true in

\section{\bf Introduction}

The importance of the symmetry (reciprocity) between the spacetime coordinate
operator ${\hat x}^\mu$ and the momentum operator ${\hat p}^\mu$ in
quantum theory was pointed out by Born~\cite{Born}. A free particle in quantum theory
is described by a wave function
\begin{equation}
\psi=\exp\biggl(\frac{i}{\hbar}p_\mu x^\mu\biggr).
\end{equation}
The wave function is completely symmetric in the two 4-vectors
$x^\mu$ and $p^\mu$. In a representation in Hilbert space of the operators
${\hat x}^\mu$ and ${\hat p}^\mu$ for which the ${\hat x}^\mu$ are
diagonal, we have
\begin{equation}
\label{poperator}
{\hat p}^\mu\rightarrow(\hbar/i)\partial/\partial x^\mu,
\end{equation}
while for diagonal ${\hat p}^\mu$, we obtain
\begin{equation}
\label{xoperator}
{\hat x}^\mu\rightarrow(-\hbar/i)\partial/\partial p^\mu.
\end{equation}
A wave function in  spacetime ($x$-space) can be Fourier transformed into another
wave function in momentum space (p-space):
\begin{equation}
\label{Fourier}
\phi(p)=\int d^4x\psi(x)\exp\biggl(\frac{i}{\hbar}p_\mu x^\mu\biggr).
\end{equation}

When we consider gravitational phenomena, we picture the universe described by a
spacetime geometry with the line element
\begin{equation}
ds^2=g_{\mu\nu}dx^\mu dx^\nu,
\end{equation}
where $g_{\mu\nu}$ is the metric tensor. In classical physics, the
momentum is describe by $mdx^\mu/d\tau$ where $m$ is the test particle mass and
$\tau$ is the proper time, and the transformation laws for $p^\mu$ are determined by
$x^\mu$. However, when we attempt to derive a quantum gravity theory, the particle
motion is not described by a geodesic, but by a wave function and a wave equation.
Following Born, we postulate a p-space line element
\begin{equation}
du^2=s_{\mu\nu}dp^\mu dp^\nu
\end{equation}
with the metric $s_{\mu\nu}$.

In analogy with the classical x-space geometry, we define the inverse
p-space metric tensor $s^{\mu\nu}$ by
\begin{equation}
s^{\lambda\nu}s_{\mu\lambda}={\delta^\nu}_\mu.
\end{equation}
Moreover, we define a p-space curvature tensor
\begin{equation}
\label{curvature}
{P^\nu}_{\alpha\lambda\mu}=\frac{\partial{L^\nu}_{\alpha\lambda}}{\partial p^\mu}
-\frac{\partial{L^\nu}_{\alpha\mu}}{\partial
p^\lambda}-{L^\nu}_{\beta\lambda}{L^\beta}_{\alpha\mu}+{L^\nu}_{\beta\mu}{L^\beta}_{\alpha\lambda},
\end{equation} where ${L^\lambda}_{\mu\nu}$ is the p-space connection
\begin{equation}
{L^\lambda}_{\mu\nu}=\frac{1}{2}s^{\lambda\sigma}\biggl(\frac{\partial
s_{\mu\sigma}}{\partial p^\nu}+\frac{\partial s_{\nu\sigma}}{\partial p^\mu}
-\frac{\partial s_{\mu\nu}}{\partial p^\sigma}\biggr).
\end{equation}
We obtain from (\ref{curvature}) the Ricci tensor
\begin{equation}
P_{\alpha\lambda}\equiv {P^\nu}_{\alpha\lambda\nu}=
\frac{\partial{L^\nu}_{\alpha\lambda}}{\partial p^\nu}
-\frac{\partial{L^\nu}_{\alpha\nu}}{\partial
p^\lambda}-{L^\nu}_{\beta\lambda}{L^\beta}_{\alpha\nu}+{L^\nu}_{\beta\nu}{L^\beta}_{\alpha\lambda}.
\end{equation}

The p-space possesses a diffeomorphism invariance in that we can define the
transformation for an arbitrary contravariant p-vector:
\begin{equation}
A^{'\mu}=\frac{\partial p^{'\mu}}{\partial p^\alpha}A^\alpha,
\end{equation}
and for a mixed p-space tensor such as ${A^\mu}_{\nu\lambda}$:
\begin{equation}
{A^{'\mu}}_{\nu\lambda}=\frac{\partial p^{'\mu}}{\partial p^\rho}\frac{\partial
p^{\sigma}}{\partial p^{'\nu}}\frac{\partial p^\tau}{\partial
p^{'\lambda}}{A^\rho}_{\sigma\tau}.
\end{equation}

Our quantum gravity theory involving two quantum geometries, identified
with the x-space and p-space geometries, leads naturally to two universal invariants,
namely, the universal invariant value of the speed of light $c$ and the invariant
maximum momentum $a$ (energy $E_M$). There have recently been interesting
proposals to obtain two such universal constants, namely, `double special relativity'
and the kinematical structure of extended special relativity, based on a Born-Infeld
electrodynamics with a hyperbolic complex structure~\cite{Amelino,Schuller}.

\section{\bf Relating Momentum Space Coordinates to Spacetime Coordinates}

In quantum mechanics and relativistic quantum field theory there is a linear relation
between spacetime and momentum space through a Fourier transform of a wave
function or a field operator in flat spacetime,
$g_{\mu\nu}=\eta_{\mu\nu}$, where $\eta_{\mu\nu}={\rm diag}(+1,-1,-1,-1)$ is the flat
spacetime metric. In the presence of a gravitational field, we lose this simple
mapping between p-space and x-space.

In quantum field theory, it is
convenient to perform calculations, such as Feynman diagrams, in the p-space
representation. For closed loop diagrams these calculations are ultraviolet
divergent, and in quantum gravity for expansions about flat space \begin{equation}
\label{expansion} g_{\mu\nu}=\eta_{\mu\nu}+h_{\mu\nu}+O(h^2), \end{equation} in
which the p-space and x-space position coordinates are related by Fourier
transforms, the theory is not renormalizable due to the dimensional nature of
Newton's constant $G$.

It seems unnatural to assume that position and momentum spaces are
independent, because this would not lead to a simple flat spacetime limit of
quantum theory or the gravitational theory based on the weak field expansion of the
metric tensor about flat spacetime.

We postulate that there exists a transformation ${\cal T}$ between x-space and
p-space with the mapping
\begin{equation}
\label{Fouriercurve}
\phi(p)=\int d^4x{\cal T}[\psi(x),x^\mu p_\mu],
\end{equation}
where ${\cal T}$ is a matrix operator. The inverse transformation must also exist
\begin{equation}
\label{Fouriercurve2}
\psi(x)=\int d^4p{\tilde{\cal T}}[\phi(p),x^\mu p_\mu].
\end{equation}
We also postulate the mapping of the metric tensor operators
\begin{equation}
{\hat g}_{\mu\nu}(x)=\int d^4p{\cal T}[{\hat s}_{\mu\nu}(p), x^\mu p_\mu],
\end{equation}
and its inverse mapping. In the flat x-space and p-space limits,
$g_{\mu\nu}=s_{\mu\nu}=\eta_{\mu\nu}$, we obtain the standard Fourier transform
(\ref{Fourier}) and its inverse transformation.

We shall also postulate that
for the transformations (\ref{Fouriercurve}) and (\ref{Fouriercurve2}) there exists a
transformation, ${\cal U}$, between x-space and p-space such that for diagonalized
${\hat x}^\mu$ in the Hilbert space of operators
\begin{equation}
{\hat p}^\mu\rightarrow{\cal U}
\biggl(\frac{\hbar}{i}\frac{\partial}{\partial
x^{\mu}}\biggr),
\end{equation}
and for diagonalized ${\hat x}^\mu$:
\begin{equation}
{\hat x}^\mu\rightarrow{\cal U}\biggl(\frac{-\hbar}{i}\frac{\partial} {\partial
p^{\mu}}\biggr).
\end{equation}
In the flat space limit these mappings reduce to the
familiar ones (\ref{poperator}) and (\ref{xoperator}).

\section{\bf Momentum Space Action Principle and Field Equations}

We choose as our p-space action
\begin{equation}
S_p=\frac{1}{2{\bar\kappa}}\int d^4p \sqrt{-s}[P+2\lambda_p]+S_c,
\end{equation}
where $P=s^{\mu\nu}P_{\mu\nu}$ is the p-space scalar curvature, $s={\rm
Det}(s_{\mu\nu})$, ${\bar\kappa}$ is a constant and $\lambda_p$ and $S_c$ are
the p-space equivalents of the x-space cosmological constant and matter action,
respectively.

The field equations obtained from the action are given by
\begin{equation}
\label{fieldequations}
P_{\mu\nu}-\frac{1}{2}s_{\mu\nu}P-\lambda_ps_{\mu\nu}=-{\bar\kappa}K_{\mu\nu},
\end{equation}
where $K_{\mu\nu}$ is the p-space equivalent of the x-space energy-momentum
stress tensor $T^{\mu\nu}$:
\begin{equation}
K_{\mu\nu}=\frac{2}{\sqrt{-s}}\biggl(\frac{\delta S_c}{\delta s^{\mu\nu}}\biggr).
\end{equation}
It satisfies the identities
\begin{equation}
\nabla_{p\nu}K^{\mu\nu}=0.
\end{equation}
Here, $\nabla_{p\nu}$ denotes the covariant derivative with respect
to the p- space connection ${L^\lambda}_{\mu\nu}$.

The corresponding spacetime field equations are Einstein's equations
\begin{equation}
R_{\mu\nu}-\frac{1}{2}g_{\mu\nu}R-\lambda_xg_{\mu\nu}=-\kappa T_{\mu\nu},
\end{equation}
where $\kappa=8\pi G/c^4$, $\lambda_x$ denotes the x-space cosmological
constant and $T^{\mu\nu}$ satisfies the
identities
\begin{equation}
\nabla_{x\nu}T^{\mu\nu}=0,
\end{equation}
where $\nabla_{x\nu}$ denotes the covariant derivative with respect to the metric
$g_{\mu\nu}$.

\section{Flat Momentum Space and the Momentum Space Null Cone}

According to the quantum gravity reciprocity symmetry, we shall identify $K^{00}$ as
the `density' of spacetime curvature per p-space 3-volume $V_{(3p)}$:
\begin{equation}
K^{00}=X(E,{\bf p}),
\end{equation}
where $X(E,{\bf p})$ denotes the spacetime curvature density as a function of the
energy $E$ and the 3-momentum ${\bf p}$. When the p-space curvature
${P^\lambda}_{\mu\nu\rho}=0$, then the p-space is `flat' with the metric
$s_{\mu\nu}=\eta_{\mu\nu}$ and the line element
\begin{equation}
\label{flatp}
du^2=\biggl(\frac{dE}{c}\biggr)^2-(dp_1^2+dp_2^2+dp_3^2).
\end{equation}

We can now define transformations between p-space coordinates $p_i\quad (i=1,2,3)$
and energy $E$ as
\begin{equation}
\label{trans1}
E^\prime=\frac{E-wp_1}{\sqrt{1-\frac{w^2}{c^2}}},
\end{equation}
\begin{equation}
p_1^\prime=\frac{p_1-\biggl(\frac{w}{c^2}\biggr)E}{\sqrt{1-\frac{w^2}{c^2}}},
\end{equation}
\begin{equation}
p_2^\prime=p_2,
\end{equation}
\begin{equation}
\label{trans4}
p_3^\prime=p_3.
\end{equation}
Here, $w$ is the `relative speed' of the primed and unprimed p-space frames. These
transformations from  the primed to the unprimed `p-frames' leave the line element
(\ref{flatp}) unchanged: $du^{'2}=du^2$. For the p-space null cone determined by
$du^2=0$, we obtain
\begin{equation}
\frac{dE}{dp}=c,
\end{equation}
where $p=\vert{\bf p}\vert.$ The transformations (\ref{trans1})-(\ref{trans4}) form
the p-space homogeneous Lorentz group $SO_p(3,1)$.

\section{Maximally Symmetric Momentum Space Solution and a Maximum Invariant
Momentum}

We shall now assume that the p-space geometry is homogeneous and that the
tensor density $K^{\mu\nu}$ is independent of the spatial
momentum coordinates $p^i$, and that it can only depend on the energy $E$. We
further assume that the p-space is isotropic on the large scale distribution of
p-space points. The mathematical expression of this postulate is
that the 3-dimensional p-space will be a space of constant curvature:
\begin{equation}
P_{\lambda\mu\nu\rho}=C(s_{\lambda\nu}s_{\mu\rho}-s_{\lambda\rho}s_{\mu\nu}),
\end{equation}
where $C={\rm constant}$. We apply this equation to the 3-dimensional
subspace of the p-space, so we shall have
\begin{equation}   \label{Pspacetensor}
P_{iklm}=C(s_{il}s_{km}-s_{im}s_{kl}).
\end{equation}
Contracting
(\ref{Pspacetensor}) with respect to $s^{im}$ we get
\begin{equation}
\label{ConPspacetensor} P_{kl}=-2Cs_{kl}.
\end{equation}
For a 3-space,
Eq.(\ref{Pspacetensor}) is equivalent to (\ref{ConPspacetensor}), so that the line
element is spherically symmetric and each of the 3-dimensional
points can be taken as the origin of the p-space coordinate system. The
metric for constant 3-dimensional p-space curvature is
\begin{equation}
\label{3spacelineelement}
d\sigma_p^2\equiv
\gamma_{ik}dp^idp^k=\frac{dp^2}{1-\zeta\biggl(\frac{p^2}{a^2}\biggr)}
+p^2(d\chi^2+\sin^2\chi d\xi^2),
\end{equation}
where $a$ is a constant invariant momentum and $\zeta$ has
the values $+1,-1,0$. For $\zeta=+1$ the p-space is a closed space of
constant curvature. We shall choose $\zeta=+1$ so that there exists a {\it maximum}
momentum $a$ corresponding to a maximum
invariant energy $E_M=ca$. The line element (\ref{3spacelineelement}) is the p-space
equivalent of the Friedmann-Robertson-Walker line element in 3-dimensional x-space:
\begin{equation}
d\sigma_x^2=\frac{dr^2}{1-\zeta
\biggl(\frac{r^2}{{\bar r}^2}\biggr)}+r^2(d\theta^2+\sin^2\theta d\phi^2),
\end{equation}
where ${\bar r}$ is a constant.

The 3-volume of our p-space is given by
\begin{equation}
\int d^3p\sqrt{\gamma}=4\pi\int dp\frac{p^2}{\sqrt{1-\frac{p^2}{a^2}}},
\end{equation}
where $\gamma={\rm Det}(\gamma_{ik})$. This leads to the differential volume element
\begin{equation}
\label{Omegap}
d\Omega_p=\frac{p^2dpd\chi d\xi\sin\chi}{\sqrt{1-\frac{p^2}{a^2}}}.
\end{equation}

We can now introduce the p-space 4-velocity,
$w^\mu=c^2p^\mu/E$, \footnote{The Hamiltonian definition of the velocity is
$v=dE/dp$.} satisfying the condition
\begin{equation}
s_{\mu\nu}w^\mu w^\nu=1.
\end{equation}
Let us consider a fluid description of the p-space source tensor
\begin{equation}
K^{\mu\nu}=[X(E)+Z(E)]w^\mu w^\nu-Z(E)s^{\mu\nu},
\end{equation}
where $X(E)$ denotes the density of spacetime curvature per p-space 3-volume
$V_{(3p)}$ and $Z(E)$ the elasticity of space. These quantities correspond
reciprocally to the energy density $\rho$ and pressure $p$, respectively, in the
x-space representation of the perfect fluid energy-momentum tensor
\begin{equation} T^{\mu\nu}=(\rho c^2+p)u^\mu u^\nu-pg^{\mu\nu}, \end{equation}
where $u^\mu=dx^\mu/d\tau$.

We choose $w^\mu$ to satisfy the condition
\begin{equation}
w^\mu=(1,0,0,0).
\end{equation}
Then the 4-dimensional p-space metric line element has the form
\begin{equation}
du^2=\biggl(\frac{dE}{c}\biggr)^2-B^2(E)d\sigma_p^2,
\end{equation}
where $d\sigma_p^2$ is given by (\ref{3spacelineelement}) and $B(E)$ is a
scale factor that depends on the energy $E$.
The volume of 4-dimensional p-space is given by
\begin{equation}
\int d^4p\sqrt{-s}= \frac{4\pi}{c}\int
dEdp\frac{B^3(E)p^2}{\sqrt{1-\frac{p^2}{a^2}}}.
\end{equation}
There are two other maximally symmetric solutions to the p-space field equation,
namely, the maximally symmetric de Sitter and anti-de Sitter solutions.

We can also consider the spherically symmetric energy independent solution of the
p-space field equations (\ref{fieldequations}) for $K_{\mu\nu}=\lambda_p=0$:
\begin{equation}
P_{\mu\nu}=0.
\end{equation}
The line element is given by
\begin{equation}
du^2=\biggl(\frac
{dE}{c}\biggr)^2\biggl(1-\frac{2A}{p}\biggr)-\frac{dp^2}{1-\frac{2A}{p}}+p^2(d\chi^2+\sin^2\chi
d\xi^2).
\end{equation}
Here, $A$ is a constant of integration associated with the
singularity at the origin of the p-space coordinates $p=0$, corresponding to
$2GM/c^2$ in the Schwarzschild solution in x-space. The solution has the
asymptotically flat p-space boundary condition
$s_{\mu\nu}\rightarrow\eta_{\mu\nu}$. We see that the line element exhibits a
momentum space `event horizon' at $p=2A$, equivalent to the spacetime Schwarzschild
black hole event horizon at $r_s=2GM/c^2$.

\section{\bf Uncertainty Principle for Spacetime and Momentum Space Metrics}

We shall postulate that our quantum gravity theory possesses an uncertainty principle
for the x- and p-space metric operators ${\hat g}_{\mu\nu}$ and
${\hat s}_{\mu\nu}$:
\begin{equation}
\Delta {\hat g}_{\mu\nu}\Delta {\hat s}_{\rho\sigma} \geq
\frac{\hbar}{a\ell}C_{\mu\nu\rho\sigma},
\end{equation}
where $a$ denotes as before the maximum momentum, $\ell$ denotes a
{\it miminum length} and
\begin{equation}
C_{\mu\nu\rho\sigma}=\eta_{\mu\rho}\eta_{\nu\sigma}+\eta_{\mu\sigma}\eta_{\nu\rho}
-\eta_{\mu\nu}\eta_{\rho\sigma}.
\end{equation}
We then have the quantum commutation relation for the metric operators
\begin{equation}
[{\hat g}_{\mu\nu},{\hat s}_{\rho\sigma}]=i\frac{\hbar}{a\ell}
C_{\mu\nu\rho\sigma}\delta(x,p).
\end{equation}

\section{Momentum Space Wave Equation}

We must obtain a wave equation for a particle to complete the p-space dynamics. To
this end, we derive a p-space Hamiltonian formulation of gravity~\cite{Wald}. For our
p-space compact manifold with boundary $\partial M$, we require that a variation
of the metric $s_{\mu\nu}$ vanishes on $\partial M$ but its normal derivative does
not. Then we must add a surface term and the action becomes
\begin{equation}
 S_p^\prime=S_p+\frac{1}{{\bar\kappa}}\int d^3p\sqrt{{f}}F,
 \end{equation}
where $F$ is the contraction of the extrinsic p-space curvature $F_{ij}$ of the
boundary 3-surface, and $f$ is ${\rm Det}(f_{ij})$ induced on the 3-surface.

The p-space metric is now given by
\begin{equation}
du^2=\biggl(\frac{WdE}{c}\biggr)-f_{ij}\biggl[W^i\biggl(\frac{dE}{c}\biggr)+dp^i\biggr]
\biggl[W^j\biggl(\frac{dE}{c}\biggr)+dp^j\biggr],
\end{equation}
where $W$ is the
p-space lapse function and $W^i$ is the shift function. The extrinisic curvature is
\begin{equation}
F_{ij}=\frac{1}{2W}\biggl[W_{i\vert k}+W_{k\vert i}-{\dot
f}_{ij}\biggr],
\end{equation}
where $\vert$ denotes covariant derivative with
respect to $f_{ij}$ and $\dot f_{ij}$ denotes differentiation with respect to the
energy $E$.

The variable conjugate to $f_{ij}$ is
\begin{equation}
\Pi^{ij}\equiv \frac{\delta {\cal L}_p}{\delta{\dot f}_{ij}}
=\frac{\sqrt{f}(F^{ij}-f^{ij}F)}{2{\bar\kappa}},
\end{equation}
where ${\cal L}_p$ is the Lagrangian density associated with the action $S_p$.
The Hamiltonian for a closed p-space geometry is given by
\begin{equation}
H_p=\int d^3p\biggl(\Pi^{ij}{\dot f}_{ij}+\Pi^i
{\dot W}_i+\Pi\dot W-{\cal L}_p\biggr)
=\int d^3p(W{\cal H}_p+W_i{\cal H}_p^i),
\end{equation}
where
\begin{equation}
{\cal H}_p=\frac{\sqrt{f}(F_{ij}F^{ij}-F^2-P^{(3)})}{2{\bar\kappa}}
=2{\bar\kappa}Q_{ijkl}\Pi^{ij}\Pi^{kl}-\frac{\sqrt{f}P^{(3)}}{2{\bar\kappa}}.
\end{equation}
Here, $P^{(3)}$ is the 3-curvature and
\begin{equation}
Q_{ijkl}=\frac{1}{2\sqrt{f}}(f_{ik}f_{jl}+f_{il}f_{jk}
-f_{ij}f_{kl}).
\end{equation}

We have two primary constraints
\begin{equation}
\Pi\equiv \frac{\delta{\cal L}_p}{\delta {\dot W}}=0,\quad \Pi^i\equiv
\frac{\delta{\cal L}_p}{\delta{\dot W}_i}=0.
\end{equation}
Because $\delta H/\delta W=\delta H/\delta W_i=0$, we have the secondary constraints
\begin{equation}
H_p=H^i_p=0.
\end{equation}

We now define a p-space wave function $\Psi[f_{ij}]$ and obtain the wave equation
\begin{equation}
\biggl[\frac{Q_{ijkl}}{(2{\bar\kappa})^2}\frac{\delta}{\delta
f_{ij}}\frac{\delta}{\delta
f_{kl}}+\frac{\sqrt{f}(P^{(3)}-2\lambda_p)}{2{\bar\kappa}}
-{K^0}_0\biggr]\Psi[f_{ij},\phi]=0, \end{equation} where $\phi$ denotes matter
fields. This is the p-space equivalent of the Wheeler-DeWitt equation in
x-space~\cite{Wheeler}. The wave function associated with the Wheeler-DeWitt
equation in x-space does not depend on the time $t$. Equivalently, for the
reciprocity symmetry we have postulated, the wave function $\Psi[f_{ij},\phi]$ does
not depend on the energy $E$ but only on the 3-geometry $f_{ij}$ and the matter
fields $\phi$. Likewise, this means that the role of energy in the p-space geometry
is unclear.

\section{\bf Applications of Invariant Maximum Momentum}

Let us consider an electromagnetic radiation field with a vector potential
$A_\mu=({\bf A},U)$ in a small localized region of spacetime, which is
approximately flat $g_{\mu\nu}\sim\eta_{\mu\nu}$. Then, the Fourier series
representations of the scalar and vector potentials $U$ and ${\bf A}$
are
\begin{equation}
U=c\biggl(\frac{8\pi}{V_{(3x)}}\biggr)^{1/2}\sum_sQ_s(t)\cos(\Theta_s),
\end{equation}
and
\begin{equation}
{\bf A}=c\biggl(\frac{8\pi}{V_{(3x)}}\biggr)^{1/2}
\sum_sB_s(t)\sin(\Theta_s),
\end{equation}
where $V_{(3x)}$ is the x-space 3-volume and
\begin{equation}
\Theta_s=\frac{2\pi \nu_s}{c}({\bf n}_s,{\bf x})+\beta_s.
\end{equation}
Here, ${\bf n}_s$ is a unit vector giving the direction of the standing wave and
$\beta_s$ is a constant.

In our quantum gravity momentum representation there is an upper limit to the
momentum, $p=a$. We assume that in an approximately flat spacetime this
remains true. This means that for a quantum system of independent particles
the increment number density of quantum states of weight $f$ in a momentum element
$d\Omega_p$ is, according to (\ref{Omegap}), given by~\cite{Born}:
\begin{equation}
dn=\biggl(\frac{f}{h^3}\biggr)d\Omega_p=\biggl(\frac{f}{h^3}\biggr)\frac{d^3p}
{\sqrt{\biggl({1-\frac{p^2}{a^2}\biggr)}}}.
\end{equation}
Because of the square
root there is a maximum number of allowed states in a quantum system.  Thus, the
total number density of quantum states is {\it finite}:
\begin{equation}
\label{finitedensity}
n=\biggl(\frac{f}{h^3}\biggr)\int
\frac{d^3p}{\sqrt{\biggl({1-\frac{p^2}{a^2}\biggr)}}}=\biggl(\frac{4\pi
fa^3}{h^3}\biggr)\int_0^1\frac{dyy^2}{\sqrt{\biggl(1-y^2\biggr)}}=\frac{f\pi^2
a^3}{h^3}. \end{equation}

Let us consider the zero-point energy associated with a system of
oscillators. For stationary states we have
\begin{equation}
E_r=\sum_sh\nu_s(n_s+\frac{1}{2}).
\end{equation}
By using the formula (\ref{finitedensity}) with $f=2$, the zero-point vacuum energy
is given by
\begin{equation}
\rho_{\rm vac}=\sum_s\frac{1}{2}h\nu_s
=\frac{c}{2}\sum_sp_s =\frac{4\pi c}{h^3}\int_0^{a_v}
\frac{dpp^3}{\sqrt{\biggl({1-\frac{p^2}{a_v^2}\biggr)}}}
$$ $$
=\frac{4\pi ca_v^4}{h^3}\int^1_0\frac{dyy^3}{\sqrt{(1-y^2)}}=\frac{8\pi
ca_v^4}{3h^3}.
\end{equation}
We see that the zero-point vacuum density of
oscillators is finite, and its magnitude is determined by the momentum scale $a_v$.

The modified energy density of radiation takes the form
\begin{equation}
\label{radiationdensity}
{\cal E}=\int^{1/b}_0d\nu{\cal U}(\nu,T).
\end{equation}
Here, we have
\begin{equation}
{\cal U}(\nu,T)=\frac{8\pi h\nu^3}
{c^3\biggl(\exp(h\nu/kT)-1\biggr)\sqrt{(1-(\nu b)^2)}},
\end{equation}
where $p=h\nu/c$ and $b=h/ac$. Eq.(\ref{radiationdensity}) gives for $b=0 (a=\infty)$
the result
\begin{equation}
\label{StefanBoltzmann}
{\cal E}=a_BT^4,
\end{equation}
where $a_B=7.56\times 10^{-15}\,{\rm erg}\,{\rm cm}^{-3}\,K^{-4}$.
Eq.(\ref{StefanBoltzmann}) is the Stefan-Boltzmann law. We see that at very high
momentum (frequency) $p\sim a$ the Planck radiation density and the Stefan-Boltzmann
law are modified by the magnitude of the maximum momentum $a$. Such a change at high
frequencies could possibly be detected in the CMB Planck spectrum.

For the entropy density of a system of particles with temperature $T$, we obtain
\begin{equation}
{\cal S}=\biggl(\frac{8\pi}{c^3T}\biggr)\int^{1/b}_0\frac{d\nu}{\sqrt{(1-(\nu
b)^2}}\biggl\{kT\ln[(1-\exp(-h\nu/kT))^{-1}]\nu^2
$$ $$
+\frac{h\nu^3}{(\exp(h\nu/kT)-1)}\biggr\}.
\end{equation}
By making the substitution $y=b\nu$ this becomes
\begin{equation}
{\cal S}=\biggl(\frac{8\pi}{c^3T}\biggr)\int_0^1
\frac{dy}{\sqrt{(1-y^2)}}\biggl\{\biggl(\frac{kT}{b^3}\biggr)\ln[(1-\exp(-hy/bkT))^{-1}]y^2
$$ $$ +\biggl(\frac{h}{b^4}\biggr)\frac{y^3}{[\exp(hy/bkT)-1]}\biggr\}.
\end{equation} The total entropy for a system of particles is finite.

For the Coulomb energy associated with charged particles, we get
\begin{equation}
\label{Coulomb}
E_c=\biggl(\frac{h^2}{\pi V_{(3x)}}\biggr)\sum_{k,l}e_ke_lR_{kl},
\end{equation}
where $e_k$ is the charge of the kth particle and
\begin{equation}
R_{kl}=\sum_s\biggl(\frac{\cos(\Theta_{sk})\cos(\Theta_{sl})}{p_s^2}\biggr)
=\frac{V_{(3x)}}{h^3}\int\frac{d^3p}{p^2\sqrt{\biggl(1-\frac{p^2}{a^2}\biggr)}}
\cos(\Theta_k)\cos(\Theta_l).
\end{equation}
By taking the mean value of $\cos(\Theta_k)\cos(\Theta_l)$ over all
directions of propagation and phases, we find
\begin{equation}
\label{Rkl}
R_{kl}=\biggl(\frac{\pi V_{(3x)}}{h^3}\biggr)\int^a_0\int^1_{-1}\frac{dpd\mu
\cos\biggl(2\pi p\mu r_{kl}/h\biggr)}{\sqrt{\biggl(1-\frac{p^2}{a^2}}\biggr)}
$$ $$
=\frac{V_{(3x)}}{h^2r_{kl}}\int_0^a\frac{dp\sin\biggl(2\pi pr_{kl}/h\biggr)}
{p\sqrt{\biggl(1-\frac{p^2}{a^2}\biggr)}},
 \end{equation}
where $r_{kl}=r_k-r_l$ denotes the distance between the charges $e_k$ and $e_l$.

We introduce
\begin{equation}
f(q)=\frac{2}{\pi}\int_0^1 \frac{dy}{\sqrt{(1-y^2)}}\frac{\sin(yq)}{y}
=\int^q_0dzJ_0(z),
\end{equation}
where $J_0(z)$ is the Bessel function. Substituting $f(q)$ into (\ref{Rkl}) gives
\begin{equation}
\label{Rkl2}
R_{kl}=\frac{\pi V_{(3x)}}{2h^2}\frac{1}{r_{kl}}
f\biggl(2\frac{r_{kl}}{r_0}\biggr),
\end{equation}
where
\begin{equation}
r_0=\frac{h}{\pi a}.
\end{equation}

Inserting (\ref{Rkl2}) into (\ref{Coulomb}), we arrive at the modified Coulomb
energy~\cite{Born}:
\begin{equation}
E_c=\frac{1}{2}\sum_{k,l}\frac{e_ke_l}{r_{kl}}f\biggl(2\frac{r_{kl}}{r_0}\biggr).
\end{equation}
We have $f(q)\rightarrow 1$ for $q\rightarrow \infty$ and
$f(q)/q\rightarrow 1$ for $q\rightarrow 0$, so that we retain the classical Coulomb
energy for $r_{kl}\gg r_0$ and a {\it finite} self-energy of a charged particle for
$r\rightarrow 0$:
\begin{equation}
E_c=\frac{e^2}{r_0}.
\end{equation}
This result is similar to the regularization of Coulomb's law in Born-Infeld
electrodynamics~\cite{Infeld}.

The result that a maximum momentum $a$ leads to a regularization of
Coulomb's law in electrodynamics, leads one to believe that a similar regularization
of the Schwarzschild singularity in the spacetime Schwarzschild solution of
Einstein gravity could be realized in quantum gravity theory.

\section{Conclusions}

We have proposed that the symmetry between the two vector operators ${\hat p}^\mu$
and ${\hat x}^\mu$ in quantum theory is a fundamental property of nature, that should
be exploited in a quantum gravity theory. Two metric tensor operators ${\hat
g}_{\mu\nu}$ and ${\hat s}_{\mu\nu}$ associated, respectively, with the geometries
of spacetime (x-space) and momentum space (p-space) are introduced with their
respective pseudo-Riemannian geometries. Field equations for both geometries are
postulated with the x-space equations being the Einstein field equations
with an energy-momentum tensor density $T_{\mu\nu}$, while the p-space field
equations are associated with a tensor density $K_{\mu\nu}$, identified with the
density of spacetime curvature reciprocal to the energy-momentum density in
spacetime. The absence of spacetime curvature produces a flat
p-space geometry. This suggests that the existence of a curved spacetime manifold
is associated with a curvature of the momentum p-space.

By assuming a closed, homogeneous and isotropic p-space geometry, we find that the
volume of the space has a maximum invariant momentum (energy) $a$ ($E_M$),
which leads to a finite statistical number density of quantum states. This is in
contrast to the spacetime volume, which for cosmological scales can be open and
infinite. Thus, the quantum gravity momentum space geometry leads to a natural
invariant, ultraviolet cutoff, $E_M$, for {\it all particle interactions} in a closed
particle system. We apply this result to the calculation of the vacuum density
$\rho_{\rm vac}$ giving a finite value. We also find that the self-energy of a
charged particle is regularized.

The quantum gravity theory proposed leads to an uncertainty principle for the two
reciprocal metric tensor operators ${\hat g}_{\mu\nu}$ and ${\hat
s}_{\mu\nu}$, involving the maximum momentum $a$ and a mimimum length $\ell$. The
momentum representation geometry lends itself best to describing microscopic
particle physics and quantum gravity, while the spacetime representation geometry
describes the macroscopic properties of the universe. Both representation geometries
complement one another and describe the quantum and large scale properties of the
universe.

For a compact p-space the x-space at short distances will be discrete and described
by a lattice structure with a minimum length $\ell$, which we
can identify with the Planck length $\ell=\ell_{PL}$, where $\ell_{PL}=\sqrt{\hbar
G/c^3}$.\footnote{A discrete lattice structure of spacetime has been promoted as a
basis for quantum gravity~\cite{Smolin}.} In order to preserve local Lorentz
symmetry, the discrete spacetime corresponds to a noncommutative
geometry~\cite{Snyder}.  For long wave-length gravity using our spacetime metric
$g_{\mu\nu}$, we must assume a limit exists that gives us the macroscopic continuum
for distances much greater than $\ell_{PL}$.

\vskip 0.3 true in

{\bf Acknowledgments}
\vskip 0.2 true in
This work was supported by the Natural Sciences and Engineering Research Council of
Canada. I thank Michael Clayton, Martin Green and Laurent Freidel for helpful and
stimulating discussions.
\vskip 0.5 true in

\end{document}